\documentclass[twocolumn,showpacs,prl,superscriptaddress,floatfix]{revtex4}

\usepackage{color}
\usepackage{tabularx}
\usepackage{epsfig}
\usepackage{amsmath}
\usepackage{amssymb}
\usepackage{graphicx}
\usepackage{wasysym}
\usepackage{dcolumn}
\usepackage{bm}
\usepackage{subfigure}
\usepackage{psfrag}
\usepackage{subfigure}
\usepackage{times}

\begin{document}

\title{Self-Organized Criticality in Glassy Spin Systems Requires
a Diverging Number of Neighbors}

\author{Juan Carlos Andresen}
\affiliation {Theoretische Physik, ETH Zurich, CH-8093 Zurich,
Switzerland}

\author{Zheng Zhu}
\affiliation{Department of Physics and Astronomy, Texas A\&M University,
College Station, Texas 77843-4242, USA}

\author{Ruben S.~Andrist}
\affiliation {Theoretische Physik, ETH Zurich, CH-8093 Zurich,
Switzerland}

\author{Helmut G.~Katzgraber}
\affiliation{Department of Physics and Astronomy, Texas A\&M University,
College Station, Texas 77843-4242, USA}
\affiliation{Theoretische Physik, ETH Zurich, CH-8093 Zurich,
Switzerland}

\author{V.~Dobrosavljevi{\'c}}
\affiliation {Department of Physics and National High Magnetic Field
Laboratory, Florida State University, Tallahassee, Florida 32306, USA}

\author{Gergely T.~Zimanyi}
\affiliation {Department of Physics, University of California, Davis,
California 95616, USA}

\date{\today}

\begin{abstract}

We investigate the conditions required for general spin systems with
frustration and disorder to display self-organized criticality, a
property which so far has been established only for the fully-connected
infinite-range Sherrington-Kirkpatrick Ising spin-glass model
[Phys.~Rev.~Lett.~{\bf 83}, 1034 (1999)]. Here we study both avalanche
and magnetization jump distributions triggered by an external magnetic
field, as well as internal field distributions in the short-range
Edwards-Anderson Ising spin glass for various space dimensions between 2
and 8, as well as the fixed-connectivity mean-field Viana-Bray model.
Our numerical results, obtained on systems of unprecedented size,
demonstrate that self-organized criticality is recovered only in the
strict limit of a diverging number of neighbors, and is not a generic
property of spin-glass models in finite space dimensions.

\end{abstract}


\maketitle

Self-organized criticality (SOC) refers to the tendency of large
dissipative systems to drive themselves into a scale-invariant critical
state without any special parameter tuning \cite{schenk:02,comment:soc}.
These phenomena are of crucial importance because fractal objects
displaying SOC are found everywhere \cite{mandelbrot:83}, e.g., in
earthquakes, in the structure of dried-out rivers, in the meandering
of sea coasts, or in the structure of galactic clusters.  Understanding
its origin, however, represents a major unresolved puzzle because
in most equilibrium systems critical behavior featuring scale-free
(fractal) patterns is found only at isolated critical points and is
not a generic feature across phase diagrams.

Pioneering work in the 1980s provided insights into the possible origin
of SOC by identifying a few theoretical examples that display it.  The
``sandpile'' \cite{bak:87} and forest-fire models \cite{drossel:92} are
hallmark examples of dynamical systems that exhibit SOC. However, these
models feature {\em ad hoc} dynamical rules, without showing how these
can be obtained from an underlying Hamiltonian. Major questions thus
remain: Can one obtain SOC from a Hamiltonian system, beyond invasion
percolation \cite{newman:94,cieplak:94}?  Is this behavior a feature of
high-dimensional models, models with a diverging number of neighbors
and/or long-range interactions, or is it a generic property of a broad
class of systems?

Work in the 1990s offered a glint of hope. The first Hamiltonian model
displaying SOC without any parameter tuning was studied in detail
by Pazmandi {\em et al.}~\cite{pazmandi:99}: the infinite-range
fully connected Sherrington-Kirkpatrick (SK) model
\cite{sherrington:75}.  Out-of-equilibrium avalanches at zero
temperature ($T=0$) triggered by varying the magnetic field were
numerically studied along the hysteresis loop. A distinct power-law
behavior in the distribution of spin avalanches, as well as of the
magnetization jumps, was established, i.e., SOC.

The possible existence of SOC was also tested in several
finite-dimensional models, but in all these cases, at least
one parameter has to be tuned.  The best-studied such model is the
random-field Ising model where ferromagnetic Ising spins are coupled
to a random field of average strength $R$. For space dimensions $d >
2$, a critical $R_c$ exists where avalanches and magnetization jumps
show SOC; i.e., the relevant distributions assume a power-law form
\cite{sethna:93,perkovic:95,perkovic:99,kuntz:98,sethna:04}.  Similar
results were found for the random-bond Ising model \cite{vives:94},
as well as the random-anisotropy Ising model \cite{vives:01}, where
by tuning a parameter SOC, can be observed.

A recent study \cite{goncalves:08} on the efficiency of hysteretic
optimization \cite{pal:06} suggests that system-spanning avalanches
might be favored in fully connected models.  However, surprisingly, no
numerical studies have been reported to date for the ``vanilla''
Edwards-Anderson Ising spin glass (EASG) \cite{edwards:75} (Gaussian
interactions with zero mean). Recently \cite{ledoussal:10,ledoussal:12},
the possibility of SOC in the EASG for $d < \infty$ was suggested. The
work is strictly valid at equilibrium (i.e., for switches in the ground
state) and is based on droplet arguments (where a critical response is
expected for fields close to zero). These results raise the question as
to whether SOC might be present in out-of-equilibrium avalanche
simulations of the EASG, as done for the SK model \cite{pazmandi:99}.

A deeper understanding of models that exhibit SOC is thus needed.
Because the SK model is thought to be the mean-field limit of the EASG,
standard lore would suggest that the EASG may display SOC for all space
dimensions $d \ge 6$ (above the upper critical dimension $d_{\rm u}$
where mean-field behavior sets in). To understand whether mean-field
behavior suffices or long-range interactions (with and without a
diverging number of neighbors) are needed, we study field-driven
avalanches at zero temperature for the EASG in $d = 2$ -- $6$, and $8$
(with $z = 2d$ neighbors), as well as the Viana-Bray (VB) model ($d =
\infty$, $z = 6$) \cite{viana:85}. In addition, we study spin glasses on
scale-free graphs \cite{katzgraber:12} where the number of neighbors is
distributed according to a power law $\sim z^{-\lambda}$. Therefore, we
probe the system below and above the (equilibrium) upper critical
dimension, as well as for different combinations of interaction range
and number of neighbors $z$. In addition, we compare to results for the
SK model ($d = \infty$, $z = N-1$, with $N$ the number of spins). Our
results demonstrate that as long as $d < \infty$, no SOC is present in
the EASG. Furthermore, no SOC is present for the VB model ($d = \infty$
but $z = 6$ fixed) or spin glasses on scale-free graphs when the edge
degree does not diverge with the system size ($\lambda > 2$). However,
for the SK model, SOC is recovered. Our results therefore indicate that
a diverging number of neighbors is the key ingredient to obtain SOC in
glassy spin systems.

\paragraph*{Model, Observables and Algorithm.---}
\label{sec:model}

We study Ising models in $d$ space dimensions with the Hamiltonian
${\mathcal H} = -\sum_{\langle i,j\rangle}^N J_{ij} S_i \,S_j - H\sum_i
S_i$.  Here $S_i = \pm 1$ represent $N = L^d$ Ising spins on hypercubic
lattices of linear size $L$. The interactions $J_{ij}$ are drawn from a
Normal distribution with zero mean, and $H$ represents a magnetic field
that drives the avalanches.  For $d = \infty$ (SK limit
\cite{sherrington:75}), the sum is over all spins and the variance of the
interactions is chosen as $1/(N-1)$. When $d < \infty$, the model is
known as the nearest-neighbor EASG \cite{edwards:75} where the
interactions have variance $1$.  The VB model is similar to the SK
model; however, the number of neighbors is fixed to $6$. In the
scale-free graphs, the distribution of $z$ decays with a power law $\sim
z^{-\lambda}$.

The algorithm used is zero-temperature Glauber dynamics
\cite{sethna:93,perkovic:99,katzgraber:02b}.  We start by computing
the local fields for all spins: $h_i=\sum_{j} J_{ij}S_j - H$.
A spin is unstable if the stability $h_i S_i < 0$. The initial
field $H$ is selected such that $H > |h_i|$ $\forall i$. The
spins are then sorted by $h_i$ and the field $H$ reduced until the
stability of the first sorted spin crosses zero, making the spin
unstable \cite{comment:opt}. This unstable spin is flipped, then
the local fields of the other spins are recalculated, and the most
unstable spin is flipped again. The process is repeated until all
spins become stable, i.e., their stabilities are non-negative. In
most cases, the flipping of the first unstable spin triggers the
flipping of a substantial number of other spins, therefore causing
avalanches. The parameters are shown in Table \ref{tab:simparams}.

\begin{table}
\caption{
Simulation parameters: For each dimension $d$, we study $N = L^d$ spins
($d < \infty$) and average over $N_{\rm sa}$ disorder samples.  For the
SK, VB, and scale-free models ($d = \infty$), we study up to $32\,000$
spins with at least $15\,000$ disorder samples.  \label{tab:simparams}
}
{\scriptsize
\begin{tabular*}{\columnwidth}{@{\extracolsep{\fill}} c r r r}
\hline
\hline
$d$ & $L$ & $N$ & $N_{\rm sa}$ \\
\hline
$2$      & $1000$ &  $1\,000\,000$  &   $15\,000$ \\
$2$      & $2000$ &  $4\,000\,000$  &   $15\,000$ \\
$2$      & $3000$ &  $9\,000\,000$  &   $14\,880$ \\
$2$      & $4000$ & $16\,000\,000$  &   $14\,860$ \\
\hline
$3$      &  $100$ &  $1\,000\,000$  &   $15\,000$ \\
$3$      &  $150$ &  $3\,375\,000$  &   $10\,000$ \\
$3$      &  $200$ &  $8\,000\,000$  &   $12\,900$ \\
$3$      &  $250$ & $15\,625\,000$  &   $14\,250$ \\
\hline
$4$      &   $10$ &      $10\,000$  &   $15\,000$ \\
$4$      &   $20$ &     $160\,000$  &   $15\,000$ \\
$4$      &   $40$ &  $2\,560\,000$  &   $15\,000$ \\
$4$      &   $60$ & $12\,960\,000$  &   $15\,000$ \\
\hline
$6$      &    $8$ &    $262\,144$   &    $15\,000$ \\
$6$      &   $10$ &  $1\,000\,000$  &    $15\,000$ \\
$6$      &   $12$ &  $2\,985\,984$  &    $15\,000$ \\
$6$      &   $14$ &  $7\,529\,536$  &    $15\,000$ \\
$6$      &   $16$ & $16\,777\,216$  &    $15\,000$ \\
\hline
$8$      &    $4$ &     $65\,536$  &     $15\,000$ \\
$8$      &    $5$ &    $390\,625$  &     $15\,000$ \\
$8$      &    $6$ & $1\,679\,616$  &     $15\,000$ \\
$8$      &    $7$ & $5\,764\,801$  &     $14\,480$ \\
$8$      &    $8$ & $16\,777\,216$ &     $10\,200$ \\
\hline
\hline
\end{tabular*}
}
\end{table}

\begin{figure}
\includegraphics[width=0.48\columnwidth]{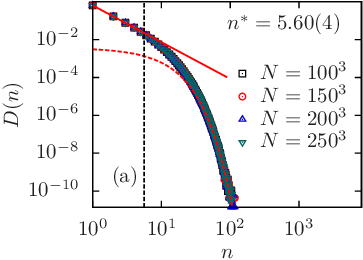}
\includegraphics[width=0.48\columnwidth]{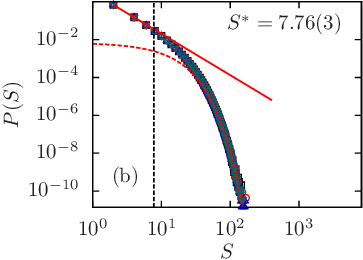}
\caption{
(a) Avalanche distribution $D(n)$ for the $d = 3$ EASG.  (b)
Magnetization jump distribution $P(S)$.  Both are recorded across the
whole hysteresis loop and the data show no finite-size effects. The
solid line represents a power law fit, whereas the dashed curve
represents an exponential fit (see the text). The vertical dashed line
marks the crossover value ($n^\ast$ and $S^\ast$ are determined by
a fit to an exponential cutoff function; see the text).
}
\label{fig:3D}
\end{figure}

At each avalanche triggered by the above algorithm, we measure the
number of spins $n$ that flipped until the system regains equilibrium
and record the distribution of avalanche sizes $D(n)$ for all triggered
avalanches until $S_i \to - S_i$ $\forall i$.  In addition, we measure
the magnetization jump $S$ at each avalanche and record the distribution
of magnetization jumps $P(S)$ \cite{comment:restr,comment:frustration}.
For the SK model, the avalanches are expected to be power law
distributed with an exponential cutoff that sets in at a characteristic
size $n^\ast$ (similar arguments are valid for the magnetization jumps
with a characteristic size $S^\ast$).  Only if $n^\ast(N) \to \infty$ as
$N \to \infty$, does the system exhibit SOC. We determine $n^\ast$ in
two different ways: First, we fit the tail of the distributions to $D(n)
\sim \exp\left[-n/n^\ast(N)\right]$ with $n^\ast(N)$ a parameter. We
also fit the small-$n$ regime to a power law and determine the point of
closest proximity between the fits.  This yields a second estimate of
$n^\ast_{\rm c}(N)$ [see Fig.~\ref{fig:3D}]. While $n^\ast(N)$ obtained
by the two approaches can differ by as much as a factor of $\sim 2$,
$n^\ast(N\to\infty)$ obtained by either definition exhibits the same
qualitative behavior. We choose to fit the distributions and extract
$n^\ast(N)$ for a given space dimension $d$ and (linearly) extrapolate
to $N = \infty$.

\begin{figure}
\includegraphics[width=0.48\columnwidth]{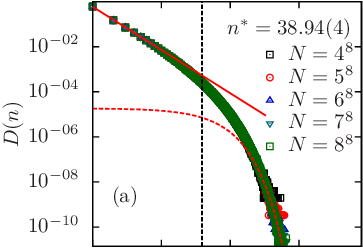}
\includegraphics[width=0.45\columnwidth]{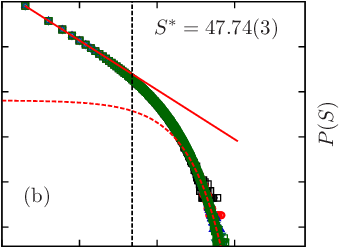}
\includegraphics[width=0.48\columnwidth]{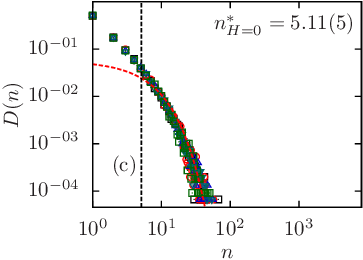}
\includegraphics[width=0.45\columnwidth]{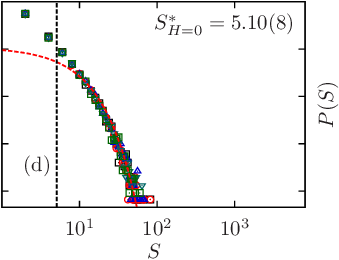}
\caption{
(a) Avalanche distribution $D(n)$ for the $d=8$ EASG. (b) Magnetization
jump distribution $P(S)$.  Both are recorded across the whole
hysteresis loop.  (c) Avalanche distribution $D_0(n)$ restricted to
$H = 0$. (d) Magnetization jump distribution $P_0(S)$ restricted
to $H = 0$.  As in Fig.~\ref{fig:3D}, the data show no finite-size
effects. The solid line represents a power-law fit, whereas the
dashed curve represents an exponential fit.  The vertical (black)
dashed lines mark the crossover value where a power law changes into
an exponential. All panels have matching vertical and horizontal scales.
}
\label{fig:8D}
\end{figure}

In addition, to study criticality for $H \sim 0$ in the short-range
systems, we measure the avalanche distribution $D_0(n)$ and
magnetization jump distribution $P_0(S)$ if and only if the field $H$
crosses zero [see Figs.~\ref{fig:8D}(c) and \ref{fig:8D}(d)]. These
measurements are necessary for short-range systems because the existence
of a spin-glass state in a field remains controversial
\cite{almeida:78,young:04,katzgraber:05c,joerg:08a,katzgraber:09b,banos:12}.
Therefore, under this restriction, we expect to probe an actual
(nonequilibrium) spin-glass state.

Reference \cite{pazmandi:99} argued that a true SOC system suppresses
avalanche formation and stabilizes itself by developing a power-law
pseudogap in $P(h)$, the distribution of stabilities, similar to the
Efros-Shklovski gap of the Coulomb glass. Requiring the system to be
stable against avalanches gives stringent bounds on the exponent and the
coefficient of the power-law form. Therefore, we also study the
distribution of local fields (stabilities) $P(h)$ for $h$ close to zero.

\paragraph*{Results.---}
\label{sec:results}

Figure \ref{fig:3D} shows the avalanche size and magnetization
jump distributions for the $d = 3$ EASG.  Avalanches remains small;
i.e., the cutoffs $n^\ast$ and $S^\ast$ do not scale with the system
size. In fact, even though we simulate over $10^7$ spins, the largest
avalanches (which occur extremely rarely) are only of approximately
100 spins. A crossover from a power law to an exponential cutoff in
the distributions occurs for rather small $n$ and $S$, respectively,
suggesting no SOC. The vertical dashed line in Fig.~\ref{fig:3D}
corresponds to the extrapolated values of $n^\ast$ and $S^\ast$,
respectively.

Reference \cite{pazmandi:99} reported that the SK model (the mean-field
limit or formulation of the EASG) exhibits SOC. Therefore, one could
expect that the EASG exhibits SOC above $d_{\rm u} = 6$. To test this
expectation, we simulate systems in $d = 8$ dimensions [see
Fig.~\ref{fig:8D}].  Again, no visible power-law behavior is present,
indicating that the system displays no SOC. To sidestep the debate over
the existence of a spin-glass state in a field, we measure the
avalanches only when $H$ crosses zero, i.e., where it is most probable
that the system is in a spin-glass state, even in a nonequilibrium type
spin-glass state \cite{comment:restr}.  Because fluctuations are large
when the restricted magnetization is measured, the data are noisy.
However, again, no signs of SOC [see Figs.~\ref{fig:8D}(c) and
\ref{fig:8D}(d)].

\begin{figure}
\includegraphics[width=0.48\columnwidth]{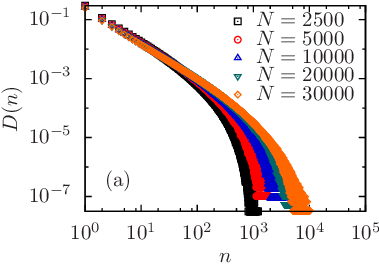}
\includegraphics[width=0.48\columnwidth]{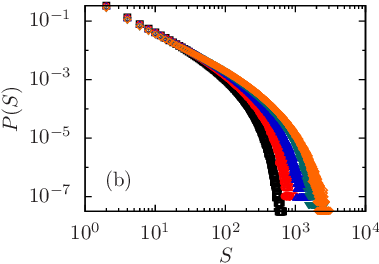}
\caption{
(a) Avalanche distribution $D(n)$ for the SK model.  (b) Magnetization
jump distribution $P(S)$.  Both are recorded across the whole
hysteresis loop. The crossover from power law to an exponential
cutoff behavior grows noticeably with $N$, signaling that the system
displays SOC.
}\label{fig:SK}
\end{figure}

Figure \ref{fig:SK} shows data for the SK model. The data are in
agreement with Ref.~\cite{pazmandi:99}: $D(n)$ [$P(S)$] has a power-law
behavior for small $n$ [$S$] with a crossover size $n^\ast(N)$
[$S^\ast(N)$] that diverges with $N$. A scaling collapse of the data
agrees with the estimates of Ref.~\cite{pazmandi:99}. Furthermore, we
find that $1/n^\ast = 0.00011(8)$ compatible with zero for $N \to
\infty$, i.e., $n^\ast = \infty$.  Data for the VB model (not shown)
show no signs of SOC and are qualitatively similar to the data shown in
Fig.~\ref{fig:3D}.  Data for spin glasses on scale-free graphs (not
shown) only show a power-law behavior (i.e., SOC) when the number of
neighbors diverges with the system size ($\lambda \le 2$,
Ref.~\cite{katzgraber:12}). Our results therefore show that a diverging
number of neighbors (node degree) is a necessary condition for SOC to be
present.

\begin{figure}
\includegraphics[width=0.80\columnwidth]{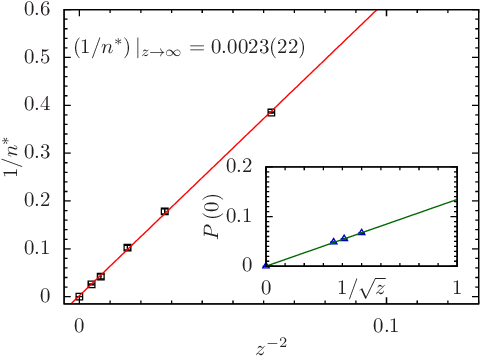}
\caption{(Color online)
Inverse of the characteristic avalanche scale $n^\ast$ as a function
of coordination $z$. The data extrapolate perfectly to the SK limit,
meaning that for {\em any} finite coordination number, the avalanches are
finite in size for the EASG. The inset shows $P(h=0)$ as a function
of $z^{-1/2}$ for different space dimensions. $P(h=0) = 0$ only for
the SK model (data from Ref.~\cite{boettcher:07a}).
}\label{fig:inset}
\end{figure}

In Fig.~\ref{fig:inset}, we plot the crossover avalanche size $n^\ast$ as
a function of the coordination $z = 2d$ and $d = 2$ -- $8$, as well as
$z = \infty$ (SK). The data show that $n^\ast \propto z^2$; i.e., a true
power-law behavior without a cutoff is only feasible for $z = \infty$
when the graph is complete \cite{comment:fit}. Note that the same
behavior is obtained for avalanches restricted to $H = 0$, as well as
the magnetization jumps. Finally, our results are independent of the
choice of $n^\ast$ (either from a fit to an exponential or from the
closest distance between the fitting function), illustrating the
robustness of the effect.  The inset of Fig.~\ref{fig:inset} shows $P(h
= 0)$ as a function of the number of neighbors $z$. The data clearly
show that $P(h = 0) \propto z^{-1/2} \to 0$ for $z \to \infty$ only;
i.e., SOC is only present for the fully connected models, such as the SK
model \cite{mezard:87,eastham:06}.

\paragraph*{Conclusions.---} 
\label{sec:conclusions}

We have demonstrated that avalanches in short-range spin glasses do not
span the system size, even if the space dimension $d$ is above the upper
critical dimension $d_{\rm u} = 6$.  Our results suggest that SOC as
found in Ref.~\cite{pazmandi:99} is not necessarily a property of the
mean-field regime but is instead a result of a diverging number of
neighbors $z$. Mean-field behavior can be reached in two equivalent
ways, either by increasing $d$ above $d_{\rm u}$ or by making the
interactions infinite ranged ($z \to \infty$).  Here, we show that these
two limits can lead to different behaviors, which become equivalent only
in the $d=\infty$ and of the infinite-range interaction limit. Analyzing
models that allow for a continuous tuning of an effective space
dimension \cite{kotliar:83,katzgraber:03,leuzzi:08} might thus also help
in gaining further insights into this problem.

One has to keep in mind, however, that the conventional arguments
\cite{mezard:87} determining $d_{\rm u}$ are restricted to equilibrium
states which below the glass transition temperature are typically
difficult or impossible to reach experimentally.  In contrast, the
metastable states that the system visits on the outer hysteresis
loop---which our avalanches explore---are indeed as far from equilibrium
as possible. One should therefore not naively and indiscriminately apply
equilibrium concepts such as the existence of the upper critical
dimension $d_{\rm u} = 6$ to the far-from-equilibrium behavior we study
here. This might explain why we do not find critical system-spanning
avalanches, as predicted in Refs.~\cite{ledoussal:10,ledoussal:12} for
static (equilibrium) response in short-range systems.

Our finding that the behavior of short-range models in any finite
dimension remains fundamentally different than that of the
fully connected infinite-range model is, therefore, a striking and a
potentially far-reaching result. It calls for a change of perspective
with respect to far-from-equilibrium states, and we hope that it will
stimulate further efforts from the theoretical and the experimental
community. The special role we suggest for the fully connected
long-range interactions may have further interesting consequences,
especially for bad metals near the metal-insulator transition
\cite{dobrosavljevic:12}. Here, the Coulomb interaction between charge
carriers assumes center stage, because poor screening in the bad metal
regime directly reveals its long-range nature
\cite{mueller:04,pankov:05,mueller:07}. Existing work has already
established that the single-particle density of states, which represents
the direct analogue of $P(h)$ in this Letter, opens a power-law
``Efros-Shklovskii'' gap within the Coulomb glass phase
\cite{mueller:04,pankov:05,mueller:07}. Given our result that the
vanishing of $P(0)$ is a direct manifestation of SOC, our findings
strongly suggest that in the presence of frustrating fully connected
long-range Coulomb interactions, SOC may survive
\cite{pastor:99,pastor:02}, even in physically-relevant space dimensions.

\begin{acknowledgments}

We would like to thank S.~Boettcher, P.~Le Doussal, M.~M\"uller, and
E.~Vives for fruitful discussions.  H.G.K.~acknowledges support from the
SNF (Grant No.~PP002-114713) and the NSF (Grant No.~DMR-1151387).
V.D.~was supported by the NSF (Grant No.~DMR-1005751).  We thank ETH
Zurich for CPU time on the Brutus cluster.

\end{acknowledgments}

\bibliography{refs,comments}

\begin{thebibliography}{47}
\bibitem[{com({\natexlab{a}})}]{comment:soc}
\bibinfo{note}{Note that within this context, SOC refers to criticality in
  zero-temperature field-driven disordered systems.}
\bibitem[{com({\natexlab{b}})}]{comment:opt}
\bibinfo{note}{We use the ``{set}'' class contained in the standard C++ library
  to store and sort the spin stabilities (local fields). We thus take advantage
  of the optimized methods included in the standard library, enabling us to
  simulate systems with over $10^7$ spins.}
\bibitem[{com({\natexlab{c}})}]{comment:restr}
\bibinfo{note}{In Ref.~\cite{pazmandi:99} the measurements are restricted to
  magnetization values in $[-0.3,0.3]$ to improve statistics. In this work this
  restriction is not necessary because of the considerably larger system sizes
  studied, as well as much larger statistics. However, to perform a fair
  comparison, we have performed the simulations with the magnetization $m$
  restricted to $[-0.3,0.3]$ obtaining qualitatively the same results.}
\bibitem[{com({\natexlab{d}})}]{comment:frustration}
\bibinfo{note}{Note that there are small differences in the results obtained
  from the distribution of avalanche sizes $D(n)$ and the distribution of
  magnetization jumps $P(S)$ due to spin flips against the field direction
  because of the inherent frustration found in spin glasses. See also
  Ref.~\cite{pazmandi:99}, where the obtained exponents for $n$ and $S$ differ
  for the same reason.}
\bibitem[{com({\natexlab{e}})}]{comment:fit}
\bibinfo{note}{We fit the data to $1/n^\ast(z) = 1/n^\ast + b/z^\gamma$.
  Allowing both $1/n^\ast$ and $\gamma$ to vary, we obtain $1/n^\ast =
  -0.0015(23)$ and $\gamma = 1.91(4)$ with $Q = 0.99$ \cite{press:95}. Fixing
  $\gamma = 2$, we obtain $1/n^\ast = 0.0023(22)$ with $Q = 0.99$. Fixing
  $1/n^\ast = 0$ and allowing $\gamma$ to vary gives $\gamma = 1.93(3)$ with $Q
  = 0.99$. Therefore, the data are compatible with $1/n^\ast = 0$ and $\gamma
  \approx 2$.}
\end{thebibliography}

\end{document}